\documentstyle[aps,epsf]{revtex}
\def\deriv #1{{#1}'}
\def\cd{{\cal D}}
\def\ie{ {i.e.,} }

\def\d{{\rm d}}

\def\dt{{\delta \theta}}

\def\dt2{(\delta \theta)^2}

\def\et{{\it et al.} }

\def\p{\sigma}
\def\al{\alpha}
\def\be{\beta}

\def\de{\delta}

\def\im{{\rm i}}

\def\lan{\left\langle}
\def\ran{\right\rangle}

\def\delx{\partial_x}

\def\nonum{\nonumber}

\def\dt{\frac{\partial}{\partial T}}

\def\e{{\rm e}}

\def\e{{\rm e }}

\def\jo #1#2#3#4{#1 {\bf #2}, #4  (#3)}  
\def\cond #1#2{#1 preprint cont-mat/#2}  

\def\PRB{Phys.\ Rev.\ B}
\def\PRL{Phys.\ Rev.\ Lett.}

\def\JPSJ{J.\ Phys.\ Soc.\ Jpn.}

\def\ADV{Adv.\ Phys.}

\def\EPL{Europhys.\ Lett.}

\def\EPJB{Eur.\ Phys.\ J.\ B}
\begin{document}
\draft
\twocolumn[    
\widetext      
\title{Transports of Disordered Carbon Nanotubes with Long Range Coulomb Interaction}
\author{Hideo Yoshioka$^{1,2}$} 
\address{
$^1$Department of Physics, Nagoya University, 
Nagoya 464-8602, Japan. \\ 
$^2$Department of Applied Physics, Delft University of Technology, 
Lorentzweg 1, 2628 CJ Delft, The Netherlands. 
}
\date{\today}
\maketitle
\vspace{-1.0cm}             
\widetext                  
\begin{abstract} 
\leftskip 54.8pt           
\rightskip 54.8pt          

Transport properties of disordered carbon nanotubes 
are investigated with including long range Coulomb interactions. 
The resistivity and optical conductivity  
are calculated by using the memory functional method. 
In addition, the effect of localization is taken into account 
by use of the renormalization group analysis and  
it is shown that the backward scattering 
of the intra-valley and that of the inter-valley cannot coexist in the localized regime.  

\end{abstract}


]                

\narrowtext
Single wall carbon nanotubes (SWNTs) are the new materials 
which are an experimental realization of one-dimensional (1-D) 
electron systems\cite{Thess}. 
Since the SWNT is made by rolling up a graphite sheet,
it is expected that fascinating properties different 
from the conventional quantum wires  made from semiconductor heterostructures
will be observed. 
From this point of view,  
transport properties of disordered SWNTs have been discussed 
in Refs.\cite{Ando-Nakanishi} and \cite{Ando-Nakanishi-Saito}. 
It has been predicted that 
the backward scattering due to the impurities vanishes 
when the range of the impurity potential 
is much larger than the lattice constant, but  
that the backward scattering reappears
when applying a magnetic field perpendicular to the tube axis.
It should be noted that the absence of the backward scattering 
holds for the graphite sheet as well as any SWNTs.         

In the above studies, 
electron correlations have been neglected. 
The 1-D nature together with the electron-electron interaction 
has been known to result in a variety of 
correlation effects in SWNTs in case of short range 
interactions\cite{Balents-Fisher,Krotov-Lee-Louie,Lin} 
and for the long range 
Coulomb interaction\cite{Egger-Gogolin,Kane-Balents-Fisher,Yoshioka-Odintsov,Odintsov-Yoshioka}.
Effects of electronic correlation in SWNTs has been measured 
in the Coulomb blockade regime\cite{Tans} as well as for Ohmic contacts\cite{Bockrath}.  
In the latter experiment, 
power-law dependences of the conductance as a function of temperature
and of the differential conductance as a function of bias voltage have been observed  
and interpreted in terms of tunneling 
into clean SWNT with the interaction. 

Even when the Coulomb interaction is taken into account, 
the conclusion of the absence of the backward scattering  
in Refs.\cite{Ando-Nakanishi} and \cite{Ando-Nakanishi-Saito} 
is not changed.  
However, 
the effects of the Coulomb interaction on the transport in disordered 
SWNTs should be observable 
in case of shorter range impurity potentials.    
In the present paper, I will discuss 
the transport properties 
of SWNTs with short range impurity potentials 
and the long range Coulomb interaction. 
It is shown that 
the interaction gives rise to an enhanced  
resistivity compared to that without the interaction.  
In addition, the interaction leads to a power-law dependence 
of the resistivity as a function of temperature 
and modifies the power of the frequency for the optical conductivity.   
In the localized regime, I find that intra-valley and inter-valley
 backward scattering  cannot coexists.  

The SWNT has metallic bands when the wrapping vector, 
${\vec w} = N_+ {\vec a}_+ + N_- {\vec a}_-$, 
satisfies the condition, $N_+ - N_- = 0$ mod 3, 
where ${\vec a}_\pm = (\pm 1, \sqrt{3})a/2$
with $a$ being the lattice constant.  
The Hamiltonian of the metallic SWNT with the long range Coulomb interaction 
is written by the slowly varying Fermi field, $\psi_{p \al s}$,  
of the sublattice $p=\pm$, the spin $s=\pm$ and 
the valley $\al = \pm$, as follows\cite{Odintsov-Yoshioka}, 
\begin{eqnarray}
{\cal H}_0 &=&  - \im v_0  \sum_{p \al s} \al \e^{-\im p \al \chi} 
\int \d x  \psi^\dagger_{p \al s} \delx \psi_{-p \al s} \nonum \\   
&+& \frac{\bar V(0)}{2} \int \d x \rho(x)^2 , 
\label{eqn:H0-1}
\end{eqnarray}
where 
$\rho(x) = \sum_{p \al s} \psi^\dagger_{p \al s} \psi_{p \al s}$, 
$v_0 \simeq 8 \times 10^5 m/s$, and 
$\bar V(0) = 2 e^2/\kappa \ln (R_s/R)$ with $\kappa \simeq 1.4$, 
$R_s$ and $R$ being the cut-off of long range of the Coulomb interaction and 
the radius of the tube, respectively.    
In Eq.(\ref{eqn:H0-1}), 
$\chi = \tan^{-1}[(N_+ - N_-)/\sqrt{3}(N_+ + N_-)]$ is the chiral angle
($\chi = 0$ corresponds to an armchair nanotube).     
In the above expression, 
I disregard the matrix elements of the interaction of the order of $a/R$, 
which lead 
to energy gaps\cite{Egger-Gogolin,Kane-Balents-Fisher,Yoshioka-Odintsov,Odintsov-Yoshioka}. 
Therefore, the present theory is valid  
when the temperature, $T$, or the frequency, $\omega$, are larger than the gaps 
induced by the Coulomb interaction. 

The impurity potential introduced as disorder of the atomic potential
is given by the following Hamiltonian\cite{Ando-Nakanishi},  
${\cal H}_{imp} = \sum_{p \al \al' s} \int \d x V^p_{\al \al'}(x) \psi^\dagger_{p \al s} \psi_{p \al' s}$, 
where $V^p_{\al \al'}(x)$ is the impurity potential at the sublattice, $p$, 
by which the electron on the valley, $\al'$, is scattered into the valley, $\al$\cite{correspondence}.  
Here I diagonalize the kinetic term in Eq.(\ref{eqn:H0-1}) and 
move to the basis of the right-moving ($r=+$) and the left-moving ($r=-$) electrons
by the unitary transformation, 
$\psi_{p \al s} = (e^{-\im p \al \chi / 2}/\sqrt{2}) \sum_r (\al r)^{(1-p)/2} \psi_{r \al s}$. 
Then, ${\cal H}_{imp}$ is written as follows, 
\begin{eqnarray}
& & {\cal H}_{imp} = \sum_{r \al s} \frac{1}{2}\int \d x 
\bigg\{
 \left[ V^+_{\al \al}(x) + V^-_{\al \al}(x) \right] \psi^\dagger_{r \al s} \psi_{r \al s} \nonum \\ 
&+& \left[ \e^{\im \al \chi} V^+_{\al -\al}(x) - \e^{- \im \al \chi} V^-_{\al -\al}(x) \right] 
\psi^\dagger_{r \al s} \psi_{r -\al s} \nonum \\
&+& \left[ V^+_{\al \al}(x) - V^-_{\al \al}(x) \right] \psi^\dagger_{r \al s} \psi_{-r \al s} \nonum \\
&+& \left[ \e^{\im \al \chi} V^+_{\al -\al}(x) + \e^{- \im \al \chi} V^-_{\al -\al}(x) \right] 
\psi^\dagger_{r \al s} \psi_{-r -\al s}
\bigg\}. 
\label{eqn:Himp-2} 
\end{eqnarray}
Here the first (second) term expresses the intra-valley (inter-valley) forward scattering, 
and the third (fourth) one is the intra-valley (inter-valley) backward scattering. 
When the range of the impurity potential is much larger than 
the lattice constant, 
$V^+_{\al \al} = V^-_{\al \al}$ and $V^+_{\al -\al} = V^-_{\al -\al} = 0$,
which leads to vanishing of the third and fourth terms in Eq.(\ref{eqn:Himp-2}), 
\ie the absence of backward scattering, 
in agreement with Ref.\cite{Ando-Nakanishi}.  
Here I consider the case of an impurity potential with range shorter than the lattice constant 
by retaining finite matrix elements of the backward scattering  
in Eq.(\ref{eqn:Himp-2}).  
I disregard forward scattering
because it does not contribute to transport. 

As was pointed out by Abrikosov and Ryzhkin\cite{Abrikosov-Ryzhkin}, 
in 1-D systems and in the limit of weak impurity potentials, 
the interaction between the electrons and the impurities 
can be parameterized by uncorrelated Gaussian random fields.  
Here I extend this method to the present model 
and introduce two kinds of the random fields, 
$\eta(x)$ and $\xi(x)$,  
expressing the intra-valley and the inter-valley backward scattering, respectively.  
The fields have the probability distributions, 
$P_\eta = \exp\left\{-(2D_1)^{-1} \int \d x \eta^2(x) \right\}$
and
$P_\xi = \exp\left\{-(D_2)^{-1} \int \d x \xi(x) \xi^*(x) \right\}$ 
where $D_1$ and $D_2$ are given by $v_0/\tau_1$ and $v_0/\tau_2$ with 
$\tau_1$ ($\tau_2$) being the scattering time due to the intra-valley (inter-valley) 
backward scattering.     
The Hamiltonian of the impurity potential is given by  
\begin{eqnarray}
{\cal H}_{imp} &=& \int \d x \eta(x) \sum_{r \al s} \psi^\dagger_{r \al s} \psi_{-r \al s} \nonum \\ 
               &+& \int \d x 
\left\{
\xi(x) \sum_{r s} \psi^\dagger_{r + s} \psi_{-r - s} + h.c.
\right\} .
\label{eqn:Himp-3}
\end{eqnarray}
Note that the intra-valley (inter-valley) backward scattering, where  
the momentum transfer in the scattering process is small (large), 
is parameterized by a real (complex) field.   
 
Here I utilize the bosonization method 
and introduce the phase variables expressing the symmetric ($\delta=+$)
and antisymmetric ($\delta=-$) modes of 
the charge ($j = \rho$) and spin ($j = \sigma$) excitations, 
$\theta_{j \de}$ and $\phi_{j \de}$.  
The phase fields satisfy the commutation
relation, $[\theta _{j\delta }(x),\phi _{j^{\prime }\delta ^{\prime
}}(x^{\prime })]={\rm i}(\pi /2){\rm sign}(x-x^{\prime })\delta _{jj^{\prime
}}\delta _{\delta \delta ^{\prime }}$. 
The Fermi field, $\psi _{r\alpha s}$, is expressed as  
\begin{equation}
\psi _{r\alpha s}=\frac{\eta _{r \alpha s}}{\sqrt{2\pi a}}
\exp \left[ {\rm i}rq_{F}x+\frac{{\rm i}r}{2} \left\{ \theta _{\alpha s}+r\phi _{\alpha
s}\right\} \right] ,   
\label{psi}
\end{equation}
where 
$\theta _{\alpha s}=\theta _{\rho+}+s\theta _{\sigma +}+\alpha \theta _{\rho -}+\alpha s\theta _{\sigma -}$
and $\phi _{\alpha s}=\phi _{\rho +}+s\phi _{\sigma +}+\alpha \phi _{\rho-}+\alpha s\phi _{\sigma -}$.
Klein factors $\eta _{r \alpha s}$ are introduced to ensure 
correct anticommutation rules for different
species $r,\alpha ,s$, and satisfy $[\eta _{r\alpha s},\eta
_{r^{\prime }\alpha ^{\prime }s^{\prime }}]_{+}=2\delta _{rr^{\prime
}}\delta _{\alpha \alpha ^{\prime }}\delta _{ss^{\prime }}$. 
The spin-conserving products $\eta _{r\alpha s}\eta _{r^{\prime }\alpha ^{\prime
}s}$ in the Hamiltonian can be represented as \cite{Egger-Gogolin}%
, $A_{++}(r,\alpha ,s)=\eta _{r\alpha s}\eta _{r\alpha s}=1$, $%
A_{+-}(r,\alpha ,s)=\eta _{r\alpha s}\eta _{r-\alpha s}={\rm i}\alpha \sigma
_{x}$, $A_{-+}(r,\alpha ,s)=\eta _{r\alpha s}\eta _{-r\alpha s}={\rm i}%
r\alpha \sigma _{z}$, and $A_{--}(r,\alpha ,s)=\eta _{r\alpha s}\eta
_{-r-\alpha s}=-{\rm i}r\sigma _{y}$ with the standard Pauli matrices $%
\sigma _{i}$ ($i=x,y,z$). 
The quantity $q_{F}=\pi n/4$ is related to the
deviation $n$ of the average electron density from half-filling and 
can be controlled by the gate voltage.  
The Hamiltonian, ${\cal H}_0$ and ${\cal H}_{imp}=\sum_{i=1,2}{\cal H}^{i}_{imp} $, 
is expressed by the phase variables as follows, 
\begin{eqnarray} 
{\cal H}_0 &=& \sum_{j=\rho ,\sigma} \sum_{\delta =\pm} 
\frac{v_{j\delta }}{2\pi }
\int {\rm d}x 
\left\{ 
K_{j\delta }^{-1}(\partial _{x}\theta _{j\delta})^{2} 
+ K_{j\delta }(\partial _{x}\phi _{j\delta })^{2}
\right\}, 
\nonum \\
& & \\
\label{eqn:H0-b}
{\cal H}^{1}_{imp} 
&=& \frac{\im \sigma_z}{2 \pi a} \int \d x \eta(x) 
\sum_{r \al s} r \al \exp\left(-2 \im r q_F x \right) \nonum \\
&\times& 
\exp\left\{
- \im r (\theta_{\rho+}+s\theta_{\p+}+\al\theta_{\rho-}+\al s \theta_{\p-})
\right\},  \\
\label{eqn:Himp1-b}
{\cal H}^{2}_{imp} 
&=&  \frac{- \im \sigma_y}{2 \pi a} \int \d x 
\sum_{r s} r \exp
\left\{- \im r ( 2 q_F x + \theta_{\rho+}+s\theta_{\p+} ) \right\} 
\nonum \\
&\times&
\left[
\xi(x) \exp \left\{- \im (\phi_{\rho-}+ s \phi_{\p-}) \right\}
+ h.c.
\right]
, 
\label{eqn:Himp2-b}
\end{eqnarray}
where $K_{\rho+} = (v_{\rho+}/v_0)^{-1} = 1/\sqrt{1+4{\bar V}(0)/(\pi v_0)}$
and $K_{j\de} = v_{j\de}/v_0 = 1$ for the others. 
The Pauli matrices seen in ${\cal H}^i_{imp}$ are due to the product of Klein factors.

With the above phase Hamiltonian, 
I calculate the dynamical conductivity, $\sigma (\omega)$,  
which is expressed by the memory function, $M(\omega)$, as follows\cite{Gotze-Wolfle},
\begin{eqnarray}
\p(\omega) &=&  \frac{- \im \chi (0)}{\omega + M(\omega)} , \\ 
\label{eqn:s-omega}
M(\omega) &=& 
\frac{\left( \langle F ; F \rangle_\omega - \langle F ; F \rangle_{\omega=0} \right)/\omega}
{- \chi (0)} ,  
\label{eqn:M-omega-1}
\end{eqnarray}
where $\langle A ; A \rangle_\omega \equiv -\im \int \d x \int_0^\infty \d t 
\e^{(\im \omega - \eta)t} \lan [A(x,t),A(0,0)] \ran$ 
with $\eta \to +0$, $\lan \cdots \ran$ denotes the thermal average 
with respect to ${\cal H}$, $F = [j,{\cal H}]$ with $j$ being the current operator,    
and $\chi (0) =  \langle j ; j \rangle_{\omega = 0}$.  
Since the present Hamiltonian conserves the total particle number, 
there are no non-linear terms including $\phi_{\rho+}$. 
Then the current operator is expressed by 
$j = 2 v_{\rho+} K_{\rho+} \delx \phi_{\rho+}/\pi$, 
which leads to $\chi (0) = - 4 v_{\rho+} K_{\rho +}/\pi$. 
To lowest order in $D_1$ and $D_2$,  
$M(\omega)$ is calculated  as 
\begin{eqnarray}
& & M(\omega) = \frac{2 \pi v_{\rho+} K_{\rho+}}{\omega} 
                   \sum_{i=1,2} \frac{D_i}{(\pi a)^2} 
               \sin \frac{\pi K_i}{2} \left( \frac{2\pi T}{\omega_F} \right)^{K_i} \frac{1}{\pi T} \nonum \\ 
         &\times&  
                   \left\{
                   B(\frac{K_i}{2} - \im \frac{\omega}{2 \pi T}, 1-K_i) - B(\frac{K_i}{2}, 1-K_i)
                   \right\}, 
\label{eqn:M-omega-2}
\end{eqnarray} 
where $K_1 = (K_{\rho+}+K_{\p+}+K_{\rho-}+K_{\p-})/2$, 
$K_2 = (K_{\rho+}+K_{\p+}+K^{-1}_{\rho-}+K^{-1}_{\p-})/2$, 
$\omega_F $ is the high energy cut-off, and 
$B(x,y) = \Gamma(x) \Gamma(y)/\Gamma(x+y)$ where 
$\Gamma(x)$ is the gamma function.  

For $\omega \to 0$, $M(\omega)$ reduces to  
\begin{eqnarray}
& & M(0) = \im 2 v_{\rho +} K_{\rho +}
                   \sum_{i=1,2} \frac{D_i}{v_0^2} \left(\frac{2\pi T}{\omega_F} \right)^{K_i -2}
                   \frac{\Gamma^2(K_i/2)}{\Gamma(K_i)} ,  
\label{M-omega-3}
\end{eqnarray} 
which leads to the resistivity, 
\begin{eqnarray}
\rho &=& \sigma(0)^{-1} = \frac{\pi^2}{2a}
\sum_{i=1,2} 
\cd_i\left(\frac{2\pi T}{\omega_F}\right)^{K_i-2} \frac{\Gamma^2(K_i/2)}{\Gamma(K_i)} ,
\label{eqn:rho-p1} 
\end{eqnarray}
where $\cd_i = D_i a / (\pi v_0^2) = a /(\pi \ell_i)$ 
with $\ell_i = v_0 \tau_i$ being the mean free path. 
In the present case of $K_i = (K_{\rho +}+3)/2$, 
the above expression results in 
$\rho = \rho_{B0} \left({2\pi T}/{\omega_F}\right)^{(K_{\rho +}-1)/2} 
{\Gamma^2((K_{\rho +}+3)/4)}/{\Gamma((K_{\rho +}+3)/2)}$ where 
$\rho_{B0} = \pi/2( \ell_1^{-1} + \ell_2^{-1})$
is the resistivity for the non-interacting system in the Born approximation. 
The above result shows that the contribution of the two kinds of backward scattering
to the resistivity is additive.  
For typical nanotubes with $K_{\rho+} \simeq 0.2$, 
the resistivity is enhanced compared to that without
the Coulomb interaction and shows a temperature dependence 
as $\rho \propto T^{(K_{\rho+}-1)/2} \simeq T^{-0.4}$\cite{power}. 
In the presence of the long range Coulomb interaction, 
the phase expressing the symmetric charge fluctuation, $\theta_{\rho+}$, 
becomes rigid compared to the noninteracting case, which is expressed by the fact of  
$K_{\rho +}$ less than unity. 
Such a rigid phase is easily pinned by the impurity potential. 
Thus, the long range Coulomb interaction enhances the resistivity. 
From the asymptotic behavior of Eq.(\ref{eqn:M-omega-2}) for $\omega \gg T$,
the optical conductivity for high frequencies is calculated as 
$\sigma (\omega) \propto \omega^{(K_{\rho +}-5)/2} \sim \omega^{-2.4}$, 
which decays faster than that for the noninteracting case, $\sigma(\omega) \propto \omega^{-2}$. 
I note that the above two results of the resistivity and the optical conductivity 
do not depend on the filling, $q_F$.  
The umklapp scattering has been known to be an other origin of resistivity.    
The umklapp scattering leads to  
$\rho \propto T^{(2K_{\rho +} -1)} \simeq T^{-0.6}$\cite{Kane-Balents-Fisher,Yoshioka-Odintsov} for 
$T \gg v_0 q_F$ and   
$\sigma(\omega) \propto \omega^{(2K_{\rho +} -3)} \simeq \omega^{-2.6}$ for 
$\omega \gg v_0 q_F$. 
Though the powers due to the umklapp scattering are similar to those given by 
the impurity scattering, 
it is possible to separate them 
by tuning the gate voltage.  

Next I take into account of  the effects of localization 
by the renormalization (RG) method. 
Following Giamarchi and Schulz\cite{Giamarchi-Schulz}, 
averaging over the random fields, $\eta(x)$ and $\xi(x)$, 
I obtain the action, $S_{imp}$, 
corresponding to the impurity potential as follows,
\begin{eqnarray}
S_{imp}^1 &=& -\frac{D_1}{2}\left(\frac{4}{\pi a}\right)^2 \int \d x \int_0^\be \d \tau \d \tau' \nonum \\
&\times& 
\Big\{
\cos (2 q_F x + \theta_{\rho+}) \cos \theta_{\p+} \sin \theta_{\rho-} \cos \theta_{\p-} \nonum \\
&\hspace{1em}& \times 
\cos (2 q_F x + \theta'_{\rho+}) \cos \theta'_{\p+} \sin \theta'_{\rho-} \cos \theta'_{\p-} \nonum \\
&+&
\sin (2 q_F x + \theta_{\rho+}) \sin \theta_{\p+} \cos \theta_{\rho-} \sin \theta_{\p-} \nonum \\
&\hspace{1em}& \times 
\sin (2 q_F x + \theta'_{\rho+}) \sin \theta'_{\p+} \cos \theta'_{\rho-} \sin \theta'_{\p-} \nonum \\
&-&2
\cos (2 q_F x + \theta_{\rho+}) \cos \theta_{\p+} \sin \theta_{\rho-} \cos \theta_{\p-} \nonum \\
&\hspace{1em}& \times 
\sin (2 q_F x + \theta'_{\rho+}) \sin \theta'_{\p+} \cos \theta'_{\rho-} \sin \theta'_{\p-} 
\Big\} , 
\label{eqn:S1}
\\
S_{imp}^2 &=& - D_2 \left(\frac{2}{\pi a}\right)^2 \int \d x \int_0^\be \d \tau \d \tau' 
\e^{-\im (\phi_{\rho-}-\phi'_{\rho-})} \nonum \\
&\times& 
\bigg\{
\sin (2 q_F x + \theta_{\rho+}) \cos \theta_{\p+} \cos \phi_{\p-} \nonum \\
&\hspace{1em}& \times 
\sin (2 q_F x + \theta'_{\rho+}) \cos \theta'_{\p+} \cos \phi'_{\p-} \nonum \\
&+&
\cos (2 q_F x + \theta_{\rho+}) \sin \theta_{\p+} \sin \phi_{\p-} \nonum \\
&\hspace{1em}& \times 
\cos (2 q_F x + \theta'_{\rho+}) \sin \theta'_{\p+} \sin \phi'_{\p-} \nonum \\
&+& \im
\Big[
\sin (2 q_F x + \theta_{\rho+}) \cos \theta_{\p+} \cos \phi_{\p-} \nonum \\
&\hspace{1em}& \times 
\cos (2 q_F x + \theta'_{\rho+}) \sin \theta'_{\p+} \sin \phi'_{\p-} \nonum \\
&-&
\cos (2 q_F x + \theta_{\rho+}) \sin \theta_{\p+} \sin \phi_{\p-} \nonum \\
&\hspace{1em}& \times 
\sin (2 q_F x + \theta'_{\rho+}) \cos \theta'_{\p+} \cos \phi'_{\p-}
\Big]
\bigg\} ,
\label{eqn:S2}
\end{eqnarray}
where 
$\theta_{j\de} = \theta_{j\de}(x,\tau)$ , 
$\theta'_{j\de} = \theta_{j\de}(x,\tau')$,
$\phi_{j\de} = \phi_{j\de}(x,\tau)$ , and  
$\phi'_{j\de} = \phi_{j\de}(x,\tau')$. 
It should be noted that 
the interaction processes generated from the equal-time component 
in Eqs.(\ref{eqn:S1}) and (\ref{eqn:S2}) are disregarded. 
The results given in the following do not qualitatively changed 
even when the such processes are included because such operators 
are less divergent than $S^i_{imp}$.    
By calculating the various correlation functions  
to the lowest order of $D_1$ and $D_2$, we have the following RG equations, 
\begin{eqnarray}
\deriv{\cd_1} &=& \left\{3 - (K_{\rho+}+K_{\p+}+K_{\rho-}+K_{\p-})/2 \right\}\cd_1 , \\
\deriv{\cd_2} &=& \left\{3 - (K_{\rho+}+K_{\p+}+K^{-1}_{\rho-}+K^{-1}_{\p-})/2 \right\}\cd_2 , \\
%
\deriv{K_{j +}} &=& - \left( {\cd_1}/{X_1} + {\cd_2}/{X_2} \right) K_{j +}^2 u_{j +} , \\
\deriv{u_{j +}} &=& - \left( {\cd_1}/{X_1} + {\cd_2}/{X_2} \right) {K_{j +}u^2_{j +}}/{16} , \\
\deriv{K_{j -}} &=& - \left( {\cd_1}K_{j -}^2 / {X_1} - {\cd_2}/{X_2}\right) u_{j -} , \\
\deriv{u_{j -}} &=& - \left( {\cd_1}K_{j -}/{X_1} + {\cd_2}K_{j -}^{-1}/{X_2} \right) {u^2_{j -}}/{16} , 
\end{eqnarray}
where $'$ denotes $d/d \ell$ with $d \ell = d \ln ({\tilde a}/a)$ 
($\tilde a$ is the new lattice constant), 
$X_1 = u_{\rho+}^{K_{\rho+}/2} u_{\p+}^{K_{\p+}/2} u_{\rho-}^{K_{\rho-}/2} u_{\p-}^{K_{\p-}/2}$,
$X_2 = u_{\rho+}^{K_{\rho+}/2} u_{\p+}^{K_{\p+}/2} u_{\rho-}^{1/2K_{\rho-}} u_{\p-}^{1/2K_{\p-}}$
and $j = \rho$ or $\p$. 
The initial conditions for the above RG equations are as follows, 
$\cd_i(0) = D_i a/(\pi v^2_0)$, 
$K_{\rho +}(0) = u_{\rho+}^{-1} = K_{\rho +}$, 
and $K_{\p +}(0) = K_{\rho -}(0) = K_{\p -}(0) 
= u_{\p +}(0) = u_{\rho -}(0) = u_{\p -}(0) = 1$.
The above equations together with the initial conditions 
indicate that $K_{\rho -} = K_{\p -}$
and $u_{\rho -} = u_{\p -}$. 
In addition, the RG equations are invariant under the transformation,
$\cd_1 \leftrightarrow \cd_2$, 
$K_{\rho -} \leftrightarrow K_{\rho -}^{-1}$, and 
$K_{\p -} \leftrightarrow K_{\p -}^{-1}$.  
Therefore I can discuss the case of $\cd_1(0) > \cd_2(0)$ 
without loss of generality. 

By solving the RG equations numerically and using 
Eq.(\ref{eqn:rho-p1}), 
the resistivity is obtained as a function of the temperature\cite{Giamarchi}.  
I show the result for $\cd_1 = 1/300$ and $\cd_2 = 1/3000$ for 
both cases with and without Coulomb interaction in Fig. 1. 
Surely, the resistivity is enhanced by the effects of the localization 
for low temperatures. 
Fig. 2 and the inset of Fig.2 show the resistivity due to 
the inter-valley scattering, $\rho_2$, and that by the intra-valley scattering, $\rho_1$,
as a function of the temperature.   
The quantity, $\rho_1$, increases monotonically with decreasing temperature, 
whereas $\rho_2$ tends to zero in the low temperature limit.  
The temperature below which a decrease of $\rho_2$ is observed 
is higher than that for the non-interacting case.    
When the inter-valley scattering is stronger 
than the intra-valley one, 
$\rho_2$ increases and $\rho_1$ vanishes. 
The result is characteristic for the localized regime 
and shows that the two mechanisms of scattering cannot coexist. 
As is seen in Eqs.(6) and (\ref{eqn:Himp2-b}), 
the intra-valley (inter-valley) scattering 
pins the phases, 
$\theta_{\rho +}$, $\theta_{\sigma +}$, $\theta_{\rho -}$, and $\theta_{\sigma -}$       
($\theta_{\rho +}$, $\theta_{\sigma +}$, $\phi_{\rho -}$, and $\phi_{\sigma -}$). 
Since the conjugate variables, $\theta_{\rho -}$ and $\phi_{\rho -}$, or  
$\theta_{\sigma -}$ and $\phi_{\sigma -}$ cannot be pinned at the same time, 
the two kinds of the backward scattering cannot coexist at low temperatures. 
In the present analysis, the quantitative discussion on the 
localized regime and crossover towards it have not done.  
Therefore, more detailed discussion are needed for understanding 
of disordered SWNTs with the Coulomb interaction.

In conclusion, 
I investigated the transport properties 
of the disordered SWNTs with Coulomb interaction.
I found that the interaction enhances the resistivity, 
leads to a power-law dependence of the resistivity as a 
function of the temperature, and modifies the power of the frequency 
of the optical conductivity. 
In addition, it is shown that the intra-valley and the inter-valley backward scattering 
cannot coexist in the localized regime. 
Finally it is mentioned that the increase of the resistance with decreasing $T$ 
observed in SWNTs\cite{Delft-1} may be due to the impurity  
scattering studied in the present paper 
because the difference of the work function of the metallic electrode 
and that of the nanotube results in a downward shift of Fermi 
level in the nanotube\cite{Delft-2}.

The author would like to thank G. E. W. Bauer, Yu. V. Nazarov, 
A. A. Odintsov, Y. Tokura and A. Khaetskii for stimulating discussions.


\begin{figure}
\centerline{\epsfxsize=6.4cm\epsfbox{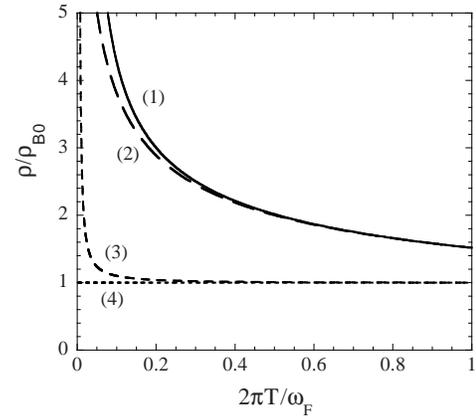}}
\caption{ 
The resistivity, $\rho$, normalized by $\rho_{B0}$ as a function of $2 \pi T/\omega_F$
in case of $\cd_1 = 1/300$ and $\cd_2 = 1/3000$. 
Here (1)((3)) is the resistivity with (without) the interaction derived from the RG
analysis, and (2)((4)) is that with (without) the interaction derived 
from the perturbation theory.
For (1) and (2), $K_{\rho +} = 0.2$ is used. 
}
\label{fig:1}
\end{figure}
\begin{figure}
\vspace{-0.8cm}
\centerline{\epsfxsize=8.0cm\epsfbox{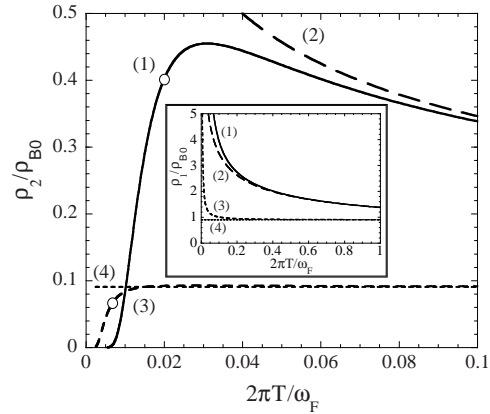}}
\vspace{-4.5cm}
\caption{ 
The normalized resistivity due to the inter-valley backward scattering, $\rho_2/\rho_{B0}$, 
as a function of $2 \pi T/\omega_F$.
The correspondence between the curves and the numbers, (1)-(4), and 
the parameters are the same as that of Fig.1.  
The white circle shows the temperature corresponding to 
$\cd_1 \simeq 1$, below which the perturbative RG analysis breaks down.  
Inset : The normalized resistivity due to the intra-valley backward scattering, $\rho_1/\rho_{B0}$,  
as a function of $2 \pi T/\omega_F$ for the same parameters. 
The correspondence between the curves and the numbers, (1)-(4), and
the parameters are the same as that of Fig.1.  
}
\label{fig:2}
\end{figure}


\begin{references}
\bibitem{correspondence}
Correspondence between the impurity potential of the present paper and 
those of Ref.\cite{Ando-Nakanishi} is as follows,  
$V^+_{\al \al} = u_A$, $V^+_{\al -\al} = u'_A$, 
$V^-_{\al \al} = u_B$, and $V^-_{\al -\al} = u'_B$.
\end{references}
\end{document}